\documentclass[aps,prd,10pt,twocolumn,superscriptaddress,nofootinbib,showkeys,showpacs,altaffilletter]{revtex4-1}

\usepackage{graphicx}
\usepackage{dcolumn}
\usepackage{amssymb}
\usepackage{amsmath}
\usepackage{amsfonts}
\usepackage{amsbsy}
\usepackage{color}
\usepackage{rotating}
\usepackage[english]{babel}

\begin{document}

\title{Recovering a redshift-extended VSL signal from galaxy surveys}

\date{\today}

\author{Vincenzo Salzano}
\email{enzo.salzano@wmf.univ.szczecin.pl}
\affiliation{Institute of Physics, University of Szczecin, Wielkopolska 15, 70-451 Szczecin, Poland}


\begin{abstract}
We investigate a new method to recover (if any) a possible varying speed of light (VSL) signal from cosmological data. It comes as an upgrade of \citep{PRL15,BAOext}, where it was argued that such signal could be detected at a single redshift location only. Here, we show how it is possible to extract information on a VSL signal on an extended redshift range. We use mock cosmological data from future galaxy surveys (BOSS, DESI, \emph{WFirst-2.4} and SKA): the sound horizon at decoupling imprinted in the clustering of galaxies (BAO) as an angular diameter distance, and the expansion rate derived from those galaxies recognized as cosmic chronometers. We find that, given the forecast sensitivities of such surveys, a $\sim1\%$ VSL signal can be detected at $3\sigma$ confidence level in the redshift interval $z \in [0.,1.55]$. Smaller signals $(\sim0.1\%)$ will be hardly detected (even if some lower possibility for a $1\sigma$ detection is still possible). Finally, we discuss the degeneration between a VSL signal and a non-null spatial curvature; we show that, given present bounds on curvature, any signal, if detected, can be attributed to a VSL signal with a very high confidence. On the other hand, our method turns out to be useful even in the classical scenario of a constant speed of light: in this case, the signal we reconstruct can be totally ascribed to spatial curvature and, thus, we might have a method to detect a $0.01$-order curvature in the same redhift range with a very high confidence.
\end{abstract}




\maketitle

\section{Introduction}
\label{sec:introduction}

The idea that fundamental constants of physics were not properly \emph{constant}, but could instead vary with time (and possibly in space) is not a new one \citep{Weyl19,Eddington23,Weyl34,Eddington34,Dirac37,Dirac38}, but a fruitful revival has been possible only quite recently, stimulated by the progress achieved in observational cosmology (for a review, see \citep{Uzan11}). On one side, the Standard Big Bang scenario suffers some theoretical shortcomings, as the horizon and flatness problems, which are at the base of the introduction of cosmological inflation \citep{Starobinsky1979,Starobinsky1980,Guth1981,Albrecht1982,Linde1982,Linde1983}. On the other, we have attended the discovery of the accelerated expansion of our Universe \citep{Riess98,Perlmutter99} and the detection of a possible variation of the fine structure constant from quasars absorption lines \citep{Webb99,Dzuba99,Webb01,Murphy01A,Murphy01B,Murphy03,Bahcall04,Chand04,Srianand04,Levshakov06,Murphy07,Srianand07,Murphy08,Molaro08,Webb11,Agafonova11,
Berengut11,Molaro13,Rahmani14}.

We have focused our attention onto the possibility that the speed of light might change in time during the evolution of the Universe; such scenario is generally called as a Varying Speed of Light (VSL) theory. A serious theoretical approach to define in the correct way a valid VSL theory is recent, and aimed exactly at solving horizon, flatness and the acceleration problems, in a ``more natural'' way, without relying on inflationary scenarios and the cosmological constant (the main successful candidate to lead Universe accelerated expansion, but also herald of many theoretical problems, see \citep{BeyondLCDM} for a review). The most exemplificative literature on this topic includes \citep{Moffat93A,Moffat93B,Albrecht99,BarrowMagueijo99A,BarrowMagueijo99B,BarrowMagueijo99C,Barrow99,Avelino99,Clayton99,BarrowMagueijo00,Bassett00,
Clayton00,Clayton01,Clayton02,Magueijo2003,new_Moffat14,new_Moffat15}. In the very own words of some of its pioneers, VSL foundations are still far from fixed, and a lot of debate is around them \citep{Ellis05,Ellis07,Magueijo2008}. But we think it should be honestly agreed that there as many brilliant and important scientists on the pro-VSL side, and not only on the against-VSL one. Thus, studying VSL theories and implications is not a fallacy, but deserves attention.

In particular, the debate concerns breaking the Lorentz invariance which VSL theories intrinsically produce, and the correctness of discussing about variation of dimensional quantities, while the only non-controversial claims should be coming from dimensionless quantities. While the former question has some possible reliable solutions \citep{Magueijo2000}, the latter is still open. Again, we want to stress here an important point, in order to make our work judged with the right perspective: we do not want to make any claim about the VSL theoretical background. This is out of the purpose of this paper and will be postponed to future works. Here, instead, we will study whether, if there is a VSL signal, and whatever is the way it can be explained, it can be detected or not, by present or future observations.

In this context, recently, we have proposed a method to \emph{measure} the speed of light \emph{on cosmological scales} and \emph{at relatively high redshift} \citep{PRL15,BAOext} using observations from galaxy surveys. This method should overcome some of the criticisms related to the fact that the speed of light is, actually, a dimensional quantity: we can measure \emph{here and now} such speed in laboratory; we have \emph{relocated} this laboratory in the outer Universe, where observations provide us a (cosmological) ruler and a (cosmological) clock, which both can be employed to measure the speed of light.

Given that we will use these rulers and clocks in this present work, we briefly review what they are. Both of them can be measured by a galaxy survey. The ruler is the sound horizon measured at late times as it is imprinted in the clustering of galaxies at cosmological scales or, equivalently, in the Baryon Acoustic Oscillations (BAO) \citep{Peebles70,Sunyaev70,Doroshkevich78,Eisenstein98A,Eisenstein98B,Cooray01,Blake03,SeoEisenstein2003,Hu03,Eisenstein05A,Eisenstein05B,SeoEisenstein2005,
Eisenstein07,Bassett10,Weinberg13}. The sound horizon has some very important properties: it is generally considered as a standard ruler, because its size in comoving coordinates is constant in time and thus can be used to calibrate/measure cosmological distances (for alternatives, see \citep{Anselmi16}); its length can be exactly calculated from theory ($\approx 150$ Mpc in physical units; the best-precision value, measured by \emph{Planck}, is $r_{s}(z_{rec}) = 144.81 \pm 0.24$ Mpc for the baseline model \citep{PlanckCosmo}). Due to the strong correlation between photons/radiation and gas in the early times (prior to the recombination epoch), by analyzing the galaxy correlation function now, it is possible to infer a correlation length which, expressed in comoving units, corresponds exactly to the sound horizon. Generally, such ruler is both present in the tangential and radial distribution of galaxies, which can respectively be defined as
\begin{equation}
y_{t}(z) = \frac{D_{A}(z)}{r_{s}(z_{rec})} \quad \mathrm{and} \quad y_{r}(z) = \frac{c_{0}}{H(z)r_{s}(z_{rec})} \, ,
\end{equation}
where $c_{0}$ is the speed of light (generally assumed constant); $z$ is the cosmological redshift; $D_{A}$ is the angular diameter distance; $H$ is the Hubble function (expansion rate); and $r_{s}(z_{dec})$ is the sound horizon, evaluated at recombination (or dragging epoch). Actually, with present data we do not have enough strong signal to measure the two directions separately, or, at least, not at the level of accuracy which should be theoretically possible \citep{Blake11,Blake12,Kazin14,BOSS2,BOSS3,BOSS4}. This will eventually be possible with future surveys, when larger number of galaxies is available; see, for example, forecast analysis for the Square Kilometer Array (SKA)\footnote{https://www.skatelescope.org/.}, \emph{Euclid}\footnote{http://sci.esa.int/euclid/.} \citep{Laureijs09,Laureijs11,Refregier10,Amendola13}, \emph{WFIRST-2.4}\footnote{http://wfirst.gsfc.nasa.gov/.} \citep{WFIRST1}, the Baryon Oscillation Spectroscopic Survey (BOSS) \citep{BOSS1,BOSS2,BOSS3,BOSS4}, the Extended BOSS survey (eBOSS) \citep{Tinker15,Prakash15,Comparat15}, the Dark Energy Spectroscopic Instrument (DESI)\footnote{http://desi.lbl.gov/.} \citep{DESI} and the Hobby-Eberly Telescope Dark Energy Experiment (HETDEX)\footnote{http://hetdex.org/.}

Galaxy surveys are beneficial for our purposes also because they can provide us the cosmological clocks we need for our method to be implemented: a sample of the observed galaxies can be targeted as {\it cosmic chronometers} \citep{Jimenez02,Moresco11,Jimenez12,Moresco12A,Moresco12B,chronometers_2}. The key idea is to find a ``cosmological clock'', which is able to give the variation of the age of the Universe with redshift. If one has this clock, then, one simply has to measure the age difference $\Delta t$ between two redshifts separated by $\Delta z$, and calculate the derivative $d z / d t \approx \Delta z / \Delta t$. Then, this quantity can be directly related to the expansion rate (Hubble function), defined as
\begin{equation}
H(z) = -\frac{1}{1+z} \frac{dz}{dt} \; .
\end{equation}
With such a method we would have a measurement of the Hubble function free from any assumption on the nature of the metric, which normally affects, for example, the definition of cosmological distances. Actually, passively-evolving early-type galaxies (ETG) turned out to be reliable candidates to play the role of such clocks. Since the first proposal, stellar population models have been improved; a much larger number of galaxies has been observed and collected, up to a redshift $z \sim 2$; and more precise tools to calibrate the clocks have been introduced (e.g., the $4000$ {\AA} break in ETG spectra). And this scenario can still be improved using future galaxy surveys in the optical, as \emph{Euclid} and \emph{WFIRST-2.4}, which should observe at least ten times more galaxies compared to the present (and ETG, eventually).

Our works \citep{PRL15,BAOext} have been recently questioned by \citep{Cai16}, where the authors point out two possible drawbacks of our method: first,
that the speed of light is measured only at one single redshift $z_{M}$; second, that we ignore the spatial curvature contribution, which is degenerate with VSL. For what concerns the first point, it is true, but we were interested in the intrinsic novelty of the method. It is well known that the angular diameter distance has a maximum at some high redshift value, which we called \emph{maximum redshift}, $z_{M}$; we found that at the maximum, the relation
\begin{equation}\label{eq:max_DA}
D_{A}(z_{M}) \cdot H(z_{M}) = c(z_{M})
\end{equation}
holds, e.g. only at the maximum redshift, the combination of the angular diameter distance and of the expansion rate is \emph{exactly equal} to the value of the speed of light at that epoch, with a minimal number of theoretical assumptions on the cosmological background, and with no need of any information at all about how the speed of light should vary or not. It is a \emph{direct} measurement, albeit local. About the second point, we have always been aware that, among the minimal number of theoretical assumptions we needed in order to derive eq.~(\ref{eq:max_DA}), we have to assume a null spatial curvature. However, we have also discussed how a non-vanishing curvature may impact our results.

Thus, in this present work, we move a step forward: using the same cosmological rulers and clocks from \citep{PRL15,BAOext}, we build a new method (different from \citep{Cai16}) which can be employed to recover a \emph{redshift-extended} VSL signal (no more limited to $z_{M}$). We also report a detailed discussion about how a non-zero curvature can influence the application of such method in a VSL theory context. Finally, we show how can the method be also applied in a classical context, where $c_{0}$ is constant in order to measure the spatial curvature itself.

In section~(\ref{sec:methodology}) we describe the theoretical apparatus at the base of our method; in section~(\ref{sec:application}) we describe all the steps required for our method to be built in more detail; in section~(\ref{sec:results}) we discuss the results obtained from the application of our method to some mock data from future galaxy surveys; and, finally, in section~(\ref{sec:conclusions}) we sum up our results.

\section{Methodology}
\label{sec:methodology}

To start, we need the observational data which is available from future BAO galaxy surveys: the angular diameter distance $(D_{A})$ and the expansion rate $(H)$. We define as $D^{real}_{A}$ and $H^{real}$ the results of such observation, i.e. the \emph{numbers} that outcome the measurement processes. Then, we need to fix the theoretical background underlying the implementation of our method. Our work is based on two main and general assumptions: we assume a Friedmann-Robertson-Walker metric and no spatial curvature. The former is the simplest and most general assumption which agrees with data and, up to some statistical accuracy, one of the main ingredients of the nowadays accepted consensus model. The latter assumption can hide a possible degeneracy between a VSL signal and the curvature; we will discuss this point in a later section and we show that, sticking to the present observational status, it is of limited concern.

Now, let us remind the theoretical definition of the angular diameter distance,
\begin{equation}\label{eq:da_theo}
D_{A} \doteq \frac{1}{1+z} \int_{0}^{z} \frac{c(z')}{H(z')} dz' ,
\end{equation}
where $H(z)$ is the theoretical Hubble function; and $c(z)$ is the speed of light expressed as any possible function of redshift. In a standard scenario, the speed of light is constant and $c(z) = c_{0}$; in the more extended context of a VSL approach, it can be any function, unknown to us until we do not recover it from the data. For what concerns $H(z)$ instead, in principle it can be derived from the first Friedmann equation (in combination with a continuity equation) once a cosmological model is given and so contains any possible information on the cosmological background. Then, we can assume that
\begin{equation}\label{eq:da_real}
D_{A} \equiv D^{real}_{A} \doteq \frac{1}{1+z} \int_{0}^{z} \frac{c(z')}{H^{real}(z')} dz',
\end{equation}
i.e., that the theoretical $D_{A}$ function (ignoring what is on the right hand side of eq.~(\ref{eq:da_real})) is \emph{explicitly} equal to the function that can be directly obtained by observations. On the other hand, we can also assume that the \emph{unknown} theoretical $H(z)$ is \emph{explicitly} equal to the function that can be obtained by observation, $H^{real}$. Actually, this is much more than just an assumption: observations always bring signatures of the real underlying cosmological model, whose ignorance we parameterize in many ways. For example, by introducing the energy-matter dimensionless parameters (eg. $\Omega_{m}$ and $\Omega_{DE}$), or the dark energy equation of state $(w_{DE})$, and so on. We highlight a very important point of our approach, which can be inferred by eq.~(\ref{eq:da_real}): we do not need any cosmological assumption (apart from the two we have stated above) for $H(z)$, because we will directly use the output from the observations, i.e. $D^{real}_{A}(z)$, in order to calculate what we have called the \emph{real} angular diameter distance.

The main point here is that we don't know, \emph{a priori}, if the speed of light appearing in eq.~(\ref{eq:da_real}) is a constant or not.

The definition of \emph{real} is just given to stress that we derive it using \emph{real} observations, and not any theoretical cosmological model. In fact, we will have to compare $D^{real}_{A}$ with another quantity, which we will call the \emph{reconstructed} angular diameter distance: this can be defined directly using $H^{real}$, i.e.
\begin{equation}\label{eq:da_rec_0}
D^{rec}_{A} \doteq \frac{1}{1+z} \int_{0}^{z} \frac{c(z')}{H^{real}(z')} dz' \; ,
\end{equation}
where, again, $H^{real}(z)$ refers to the expansion rate function directly derived from observations, and not from a theoretical model. As clearly shown above through eqs.~(\ref{eq:da_real}) and (\ref{eq:da_rec_0}), in order to convert $H^{real}(z)$, with the dimension of the inverse of time, to a distance, we need to make some assumption on the speed of light. The most general and common assumption is that it was constant; thus, the reconstructed angular diameter distance is:
\begin{equation}\label{eq:da_rec}
D^{rec}_{A} \doteq \frac{1}{1+z} \int_{0}^{z} \frac{c_{0}}{H^{real}(z')} dz' \; ,
\end{equation}
Strictly speaking, this quantity is quite useless from a cosmological point of view. Generally, all the information we need is hidden in $H(z)$: one proposes a cosmological model, which leads to $H(z)$ as a function of some parameters; and finally one tries a fit of this model with observational data, in order to recover some information about it. Here there is no benefit in using eqs.~(\ref{eq:da_rec_0}) or (\ref{eq:da_rec}): if we use directly the numbers coming out of observations (i.e. $H^{real}$), without any underlying theoretical background, we are loosing any possibility to recover the information on the cosmological model. But, if we change our perspective, and we strictly look at VSL theories, then they reveal their benefit. In fact we face a question: what if we have a \emph{real} VSL to be detected? In this case, by comparing eqs.~(\ref{eq:da_theo})~-~(\ref{eq:da_real}) with eq.~(\ref{eq:da_rec}), we can easily check that if
\begin{equation}
D^{real}_{A}(z) = D^{rec}_{A}(z)\; ,
\end{equation}
then the assumption we made to define $D^{rec}_{A}$, i.e. $c(z)= c_{0}$, was right and, indeed, $c(z) = c_{0}$. But, if instead we find out that
\begin{equation}
D^{real}_{A}(z) \neq D^{rec}_{A}(z)\; ,
\end{equation}
then we might conclude that this same assumption is wrong, and that $c(z) \neq c_{0}$. This also means that, by comparing $D^{real}_{A}$ with $D^{rec}_{A}$, we should be able to reconstruct the real unknown function $c(z)$.

Actually, this is not really straightforward in this case because the speed of light $c(z)$ enters the definition of the angular diameter distance through an integral, thus, the VSL signal cannot be isolated so easily and will result to be smoothed. But things go much easier if we focus on $H^{real}$ instead, and a possible \emph{reconstructed} $H^{rec}$. In this case, on one side we will have the \emph{direct} observational data, obtained from the derivative of the real observed $D^{real}_{A}$ as
\begin{equation}\label{eq:h_real}
y^{real}_{r}(z) \doteq \frac{\partial}{\partial z} \left[ (1+z) D^{real}_{A}(z) \right] \equiv \frac{c(z)}{H^{real}(z)}\; ,
\end{equation}
where, again, we have identified the unknown theoretical $H(z)$ function in eq.~(\ref{eq:da_theo}) with the observed $H^{real}(z)$, as we did in eq.~(\ref{eq:da_real}). On the other, we will have a \emph{reconstructed} set of
\begin{equation}\label{eq:h_real_c}
y^{rec}_{r}(z) \doteq \frac{c_{0}}{H^{real}(z)}\; ,
\end{equation}
where we need an explicit assumption of a constant speed of light in order to convert time observations $(H)$ into distances ($y_{r}$). Again, if we find that
\begin{equation}
y^{real}_{r}(z) = y^{rec}_{r}(z)\; ,
\end{equation}
then the assumption that the speed of light is constant will reveal to be well based. On the contrary, if
\begin{equation}
y^{rec}_{r}(z) \neq y^{rec}_{r}(z)\; ,
\end{equation}
then $c(z) \neq c_{0}$. What is important to stress is that, by working with $y_{r}$, we can directly obtain (or reconstruct) the analytical redshift function $c(z)$, through the ratio:
\begin{equation}\label{eq:rec_c}
\frac{y^{real}_{r}}{y^{rec}_{r}} = \frac{c(z)}{c_{0}}\; .
\end{equation}
In this way we are also circumventing the ``dimensionless-dimensional measurement'' debate, because we are going to reconstruct a (dimensionless) relative variation of the speed of light, not an absolute (dimensional) quantity.

\section{Application}
\label{sec:application}

Given the methodological basis of our model, we will now describe how to apply it in the best possible way and, in particular, we will focus on what are the limits of the accessible information we should expect from future surveys.

\subsection{Mock data}

The first point to be addressed is: what kind of data we are going to use? As stated in the section~\ref{sec:introduction}, our purpose is to show how to employ future galaxy surveys for a non-standard cosmological analysis. From such surveys we will expect to obtain separate information on $D_{A}$ (from BAO) and $H$ (from cosmic chronometers). At the present stage, we do not have yet independent measurements on $D_{A}$ and $H$, so we will need to produce mock data for our analysis, in the style of \citep{PRL15} and \citep{BAOext}. As we have pointed out in the section~\ref{sec:introduction}, there is no uniformity in approaching a VSL scenario; but we also want to stress that the exact choice of the VSL approach is meaningless in our case. We are not going to have any fit, or any test of any particular model; we only need to produce some mock observational data with a VSL signal included. The only requirement we will ask for is that, at least, such mock data were compatible with the present observations and with the present consensus model, at least in the redshift range now covered. In such a case, clearly, the VSL model would be indistinguishable from the standard scenario, but still compatible with observations. This sounds like a quite reasonable requirement: ``observations are observations'', what we measure and see is independent of our understanding of the underlying theory. Of course, a VSL signal can imply a different physical evolution in/of some processes, but the measured outcomes cannot be different from what we see now. For example: the sound horizon can be obviously influenced by a VSL. But it can also be measured with a very high confidence in a cosmic microwave background experiment. Theory has to adjust to this measurement, not viceversa. Thus, if we create a mock VSL data set which is compatible with present observations, we are just implying that the VSL signal has to be consistent with them.

Following \citep{BarrowMagueijo99B}, if a VSL is introduced with a minimal coupling with gravity, then we have modified versions of the first Friedmann equation and of the continuity equation which are, respectively:
\begin{equation}
H^2(t) = \frac{8\pi G}{3} \rho(t) - \frac{k}{a^2(t)}c^{2}(t)\; ,
\end{equation}
and:
\begin{equation}
\dot{\rho}(t) + 3 H(t) \left( \rho(t) + \frac{p(t)}{c^2(t)}\right) = \frac{3 k}{4\pi G a^2(t)}c(t)\dot{c}(t)\; ,
\end{equation}
where: $\rho$ and $p$ are, respectively, the mass density and the pressure of any fluid in the Universe; $a(t)$ is the scale factor; $G$ is the universal gravitational constant; and the speed of light is expressed as a general function of time (or redshift), $c(t)$. As we have anticipated before, a degeneracy between VSL and geometry is possible: indeed, any change produced by a VSL is connected with the spatial curvature. We will discuss this later in more detail; for now, we will assume that Universe is spatially flat, e.g. $k=0$, which implies that no effective change is effective in the continuity equation and, consequently, in the first Friedmann equation (at least, in terms of the energy-mass equations of state). On the other hand, in the calculation of $D^{real}_{A}$, the VSL also operates through the $c(z)$ function which enters the integral.

It is thus clear that, in order to produce our mock data, we need to assume an ansatz for $c(z)$; we follow \citep{Magueijo2003} and consider the ansatz:
\begin{equation}\label{eq:ansatz_c}
c(a) \propto c_{0} \left( 1+ a/a_{c} \right)^{n}\; ,
\end{equation}
where $a \equiv 1/(1+z)$ is the scale factor, and $a_{c}$ sets the transition epoch from some $c(a) \neq c_{0}$ (at early times) to $c(a) \rightarrow c_{0}$ (now).

The fiducial cosmological model used to produce the mock data in this work is a slightly modified version of the baseline $\Lambda$CDM model from \emph{Planck} 2015 release\footnote{http://wiki.cosmos.esa.int/planckpla/index.php\\/Cosmological$\_$Parameters.
}, \texttt{base}$\_$\texttt{plikHM}$\_$\texttt{TTTEEE}$\_$\texttt{lowTEB}$\_$\texttt{lensing}$\_$\texttt{post}$\_$\texttt{BAO}. This model is characterized by a dimensionless matter density today equal to $\Omega_{m} = 0.31$. We have to modify slightly this parameter when introducing a VSL because a VSL can mimic an accelerated expansion (this was the original motivation for starting to study VSL theories) and, thus, can be seen as a contribution to the dark energy sector. More precisely, in a VSL context, the acceleration would not be given by a real cosmological fluid, but would be an implicit effect due to a varying speed of light. In any way, this means that, when adding a VSL, the contribution from a dark energy fluid diminishes and, consequently, in a spatially flat Universe, $\Omega_{m}$ might grow. In this work we have considered two different VSL scenarios: one, given by the paramaters $a_{c} = 0.05$ and $n = -0.001$, corresponds to a redshift-increasing speed of light, with an average variation $\sim 0.1\%$ at redshift $1.5-1.6$ ; the other, given by the paramaters $a_{c} = 0.05$ and $n = -0.01$, corresponds to a redshift-increasing speed of light, with an average variation $\sim 1\%$ at redshift $1.5-1.6$. This redshift range is used as a reference, following the nomenclature used in \citep{PRL15,BAOext}. In Table~I of \citep{BAOext} we also show how both the models are consistent with present observations and with the fiducial cosmological model from \emph{Planck}, if we assume for them, respectively, $\Omega_{m} = 0.314$ and $\Omega_{m} = 0.348$.

Once we have defined our input cosmological model, we can produce the mock $H^{fid}$ and $D^{fid}_{A}$; but in order to produce realistic mock data, i.e., the previously defined $H^{real}$ and $D^{real}_{A}$, we still need the observational errors on these quantities. In \citep{FontRibera2014}, many on-going and future surveys are analyzed; the authors analyze what are the errors we should expect on such quantities by each one of these surveys, assuming redshift bins of $0.1$ width. Among them, we will focus on: BOSS, DESI, and \emph{WFIRST-2.4} because, in their respective redshift ranges, they show the best performances. For BOSS, we will consider $z=0.05$; for DESI, $z \in [0.15,0.55]$; for \emph{WFIRST-2.4}, $z \in [1.95, 2.75]$. For the intermediate range $z \in [0.65,1.85]$ we use the SKA results from \citep{SKA}; the performance of SKA will outcome the others by at least one order of magnitude, thus, it will be quite natural to expect the best results of our approach from this redshift range.

With $H^{fid}$ and $D^{fid}_{A}$ and the corresponding errors, we can now simulate realistic data: we randomly generate our values of $H^{real}$ and $D^{real}_{A}$ from a multivariate Gaussian centered on the fiducial values, and with a total covariance matrix built up from the errors defined for each survey. We also additionally assume a correlation factor between $H^{fid}$ and $D^{fid}_{A}$ $\sim 0.4$, as derived in \citep{SeoEisenstein2007}.

We want to stress here two points. First: the errors from \citep{FontRibera2014} are not given directly in terms for $H$ and $D_{A}$ but, instead, for $H \cdot r_s(z_{\ast})$ and $D_{A} / r_s(z_{\ast})$, where $r_s(z_{\ast})$ is the sound horizon at the decoupling/dragging epoch. Thus, if we want to work with $H$ and $D_{A}$ derived from a BAO survey, we need to multiply the previous defined combination by the sound horizon. Examining all the cases covered by the \emph{Planck} mission and collected in the \emph{Planck} Legacy Archive\footnote{http://wiki.cosmos.esa.int/planckpla/index.php/\\Cosmological$\_$Parameters.}, it is possible to check that the dependence of the sound horizon on the cosmological model is very weak (on the other hand, if this were not the case, it could not be considered as a standard ruler). And its dispersion is much smaller than the observational error, which is $\approx 0.15 \%$. Thus, the contribution of the sound horizon to the total error budget, whatever is the value used for it, is quite negligible.

Second: we do not have this problem for $H$, because we assume it is derived from cosmic chronometers; but we lack a forecast analysis for the errors expected on $H$ using this probe for future surveys. In \citep{chronometers_2} we have some estimations from \emph{Euclid}, with a minimum statistical error $\sim 5\%$. For our analysis we will use the $H$ errors estimated by a BAO survey; we have to think about them as a possible precision goal for future surveys (next to Stage IV), but they will always give us a precise indication of how feasible and applicable is our approach.

\subsection{Fitting quantities}

Before using eq.~(\ref{eq:rec_c}) in order to reconstruct a possible VSL signal, many questions have to be addressed and problems solved. The first one is intrinsic to our definition of $y^{real}_{r}$ given by eq.~(\ref{eq:h_real}): it is the derivative with respect to redshift of a quantity $(D^{real}_{A})$ which is represented by a discrete set of points (observations) which have an intrinsic dispersion around the underlying fiducial cosmological model. The problems related to the dispersion cannot be avoided: the dispersion is intrinsic to the measurement process, and we can only hope to have, in the future, better measurements which can reduce it (but its nature is not of statistical origin only). Thus, we will always have an intrinsic systematic error in the derivation of $y^{real}_{r}$; moreover, the dispersion alters the derivative calculation and thus, as it is known and expected, the errors on the derivated quantity tend to explode.

Having assumed that this problem cannot be avoided, we can rely on another property of our approach: given that we are not interested in the explicit form of $H$, because we will directly use observations to infer a function which \emph{interpolates} them, we are not forced to fit our quantities following some cosmological-model-based requirements. Thus, we can try a fit based on the best analytic functions which can work in this situation and, eventually, we can apply general physical priors to such functions. But, how do we actually apply such priors? First, we generate a random mock data set $(H^{real},D^{real}_{A})$; then, we apply the fits we will describe in what follows in this section; if a generated set satisfies the chosen priors, we save it; and if not, we reject it. At the end, we are left with $10^3$ random \emph{physically realistic} mock data sets, and we apply our algorithm to each of them.

In our case, we need analytic functions for fitting both $H^{real}$ and $D^{real}_{A}$, and they have different requirements. For $H^{real}$ we have found that a simple sixth-order redshift polynomial gives an optimal fit to $H^{real}$ in the redshift range we are covering, i.e. $z \in [0.05, 2.75]$; higher-order polynomials do not improve the fit. We also apply a relaxed prior requiring that $H^{real}(z=0)$ obtained by such polynomial fit lies in the range $[H_{0}-3\sigma,H_{0}+3\sigma]$, where $H_{0}$ is an independent measurement of the Hubble constant, and $\sigma$ its error. In our case, we have used the value $H_{0} = 69.6 \pm 0.7$ from \citep{Hubble0}. As a further prior, we can also ask that $H(z)>0$ all over the redshift range $z \in [0,\infty)$.

For $D^{real}_{A}$ a polynomial fit is unsatisfactory to describe the peculiar property of the angular diameter distance to have a maximum at relatively low redshift values. A better and more flexible fit is given by the Pad$\mathrm{\acute{e}}$ approximant:
\begin{equation}\label{eq:pade_da}
D^{real}_{A}(z) = \frac{d^{t}_{1}\, z}{1 + d^{b}_{1}\, z + d^{b}_{2}\, z^{2}}\; ,
\end{equation}
which clearly satisfies the expected conditions: $D^{real}_{A} = 0$ for $z \rightarrow 0$ and $z \rightarrow \infty$; moreover, we require that $D^{real}_{A} > 0$ for $z \in [0,\infty)$, and that $(\partial D^{real}_{A}/\partial z)/c_{0}$ lies in the range $[H_{0}-3\sigma,H_{0}+3\sigma]$.

Once the fits are run for both $H^{real}$ and $D^{real}_{A}$, we have a set of parameters (the parameters of the polynomial and of the Pad$\mathrm{\acute{e}}$ approximant), respectively, with their covariance matrix and errors bars; after the correct propagation error rules are applied, we end up with a set of polynomial-reconstructed $y^{real}_{r}$ and $y^{rec}_{r}$, with related errors, from which we can derive the $c(z)/c_{0}$ ratio through eq.~(\ref{eq:rec_c}). Finally, this last quantity can also be fitted (or reconstructed); the function we have been working with is the Pad$\mathrm{\acute{e}}$ approximant given by:
\begin{equation}\label{eq:pade_cz}
\frac{c(z)}{c_{0}} = \frac{1 + c^{t}_{1}\, z}{1 + c^{b}_{1}\, z + c^{b}_{2}\, z^{2}}\; ,
\end{equation}
imposing the conditions: $c(z=0)/c_{0} = 1$, and that $c(z)$ is always positive for $z \in [0,\infty)$.

We have verified that the functions we have finally chosen to fit $H^{real}$ and $D^{real}_{A}$ are really good approximations to the fiducial model all over the entire redshift range $z \in [0,\infty)$, and not only in the redshift interval we have decided to work with because covered by next galaxy surveys. On the other hand, the function chosen for $c(z)$ has some degree of arbitrariness: it describes very well our input VSL in the galaxy surveys redshift range, but not at very high redshifts. But it is very general, and with such a high level of flexibility that it can be used as a testing function to detect if a VSL signal is working or not in any case (but any suggestion can be considered).

\section{Results}
\label{sec:results}

The first point to be examined is how good is the reconstruction of the hidden VSL signal; then, we will move to a much detailed analysis of the possible degeneracy between the VSL signal and a non-zero spatial curvature.

\subsection{Pure VSL signal}

\begin{figure*}[htbp]
\centering
  \includegraphics[width=0.9\textwidth]{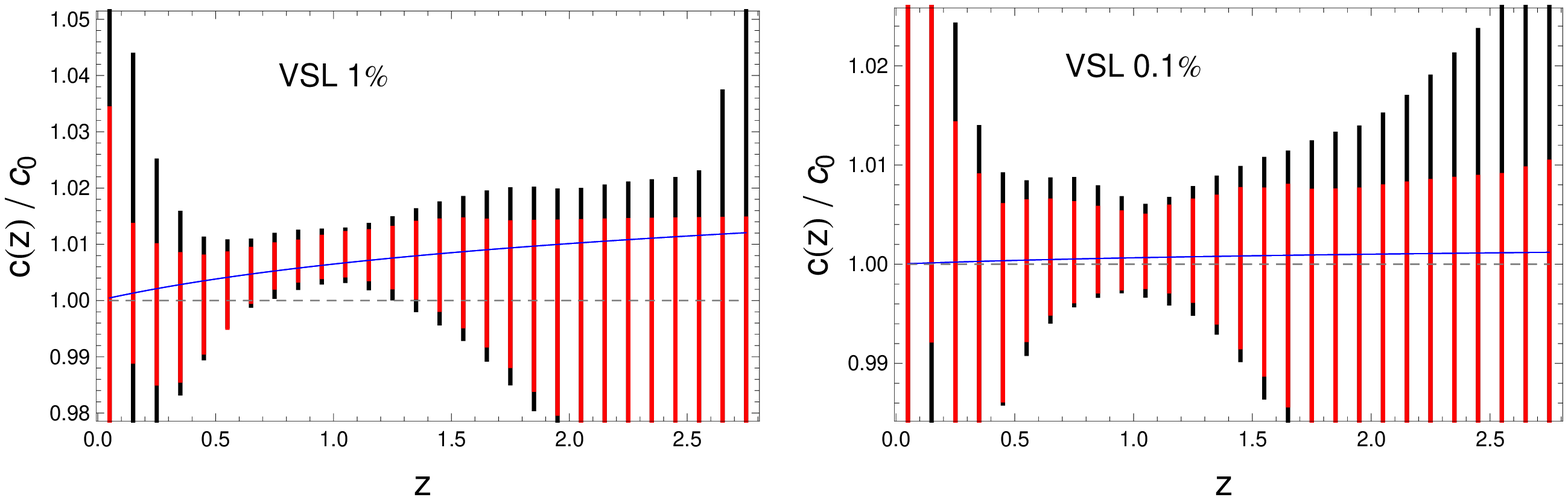}
  \caption{Reconstruction of $c(z)$ from mock data: results. Black: $1\sigma$ confidence level from the total $10^3$ simulations for eq.~(\ref{eq:rec_c}); red: $1\sigma$ confidence level from the total $10^3$ simulations for eq.~(\ref{eq:rec_c}) after the ratio $c(z)/c_{0}$ is fit with eq.~(\ref{eq:rec_c}); blue: VSL $c(z)/c_{0}$ used as input from eq.~(\ref{eq:ansatz_c}); dashed grey line: standard constant $c(z)/c_{0} = 1$. Different ranges are shown on the vertical axis. \label{fig:pend}}
\end{figure*}

In our case, we know what is the behavior of $c(z)$, so that we can easily check if the final reconstructed $c(z)$ gives a reliable description of this known input, thus testing if our algorithm works well or not. In Figure~\ref{fig:pend} we show both the cases of a $1\%$ and of a $0.1\%$ VSL signal, joining results from all the $10^{3}$ simulations we have realized. In black, we show the $1\sigma$ confidence level span by all the simulations for the VSL pre-fitting signal in eq.~(\ref{eq:rec_c}), i.e. the quantity $c(z)/c_{0}$ calculated from $y^{real}_{r}$ and $y^{rec}_{r}$ after fitting $H^{real}$ with a sixth-order polynomial and $D^{real}_{A}$ with the Pad$\mathrm{\acute{e}}$ approximant given in eq.~(\ref{eq:pade_da}). In red, we plot the $1\sigma$ confidence level span by all the simulations for the VSL signal in eq.~(\ref{eq:rec_c}) after fitting the ratio $y^{real}_{r}/y^{rec}_{r}$ with the Pad$\mathrm{\acute{e}}$ approximant given in eq.~(\ref{eq:pade_cz}).

From a simple visual inspection, it can be seen that, at least at $1\sigma$ level, a $1\%$ VSL signal can in principle be detected in the redshift range $[0.75,1.25]$, where the angular diameter distance from BAO and the Hubble function from cosmic chronometers have been assumed to have the precision actually forecast from SKA. On the other hand, it is also clear that a $0.1\%$ VSL signal will be hardly detected with the same prescriptions.

\begin{figure*}[htbp]
\centering
  \includegraphics[width=0.9\textwidth]{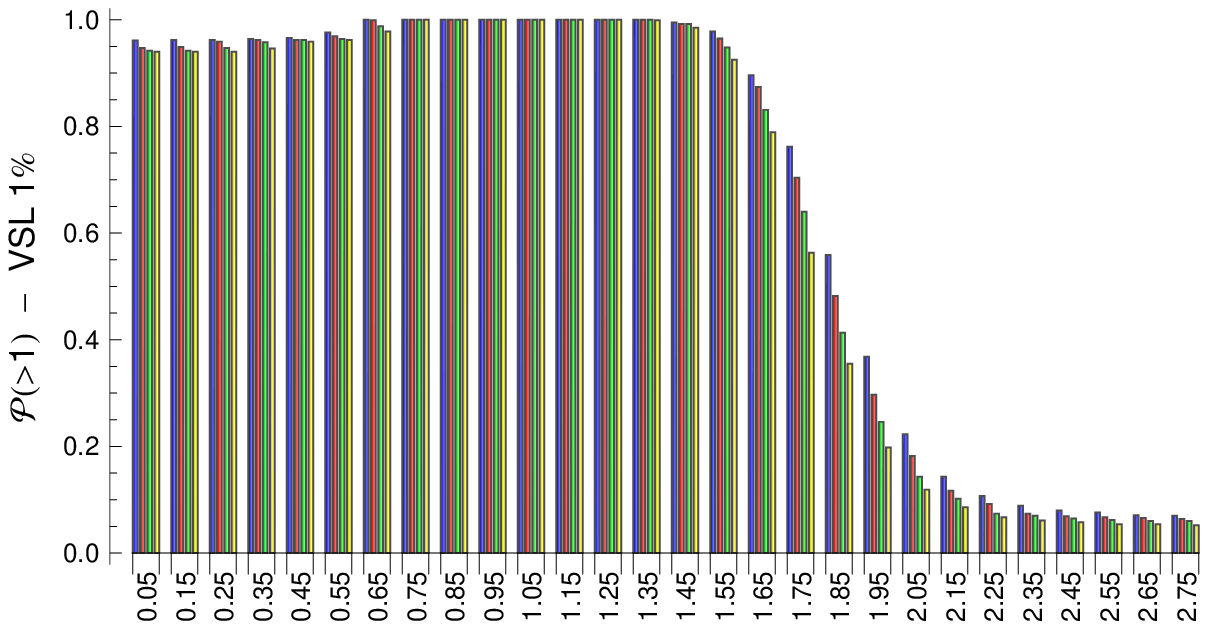}\\
  ~~~\\
  \includegraphics[width=0.9\textwidth]{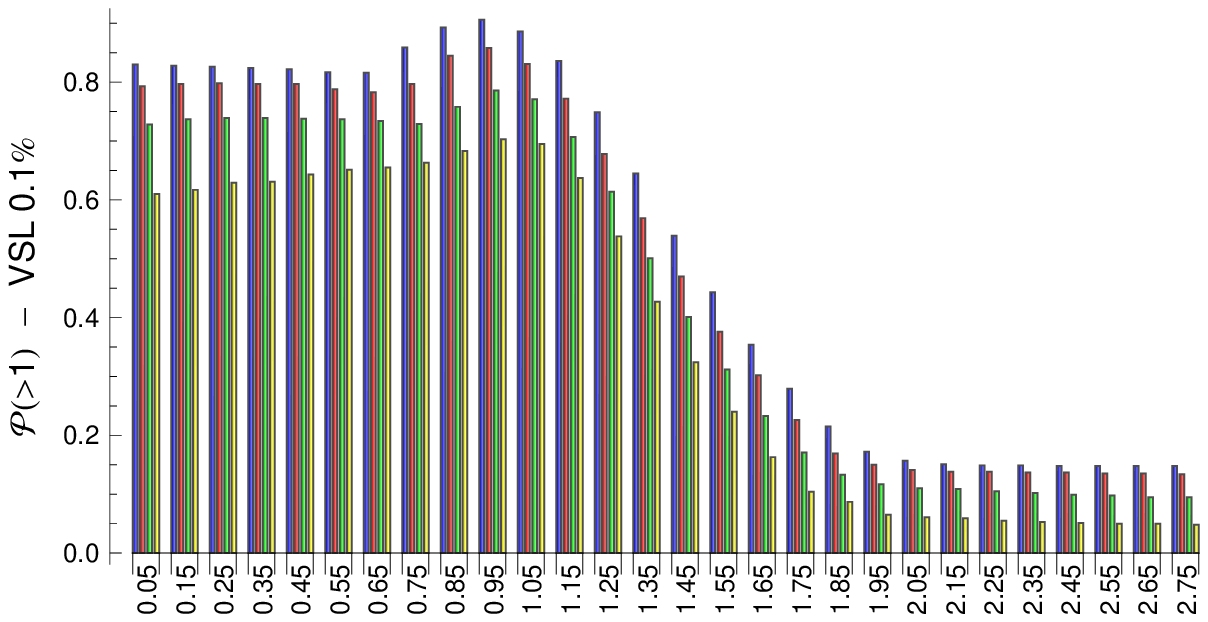}
  \caption{Probability to detect positive residuals of post-fitting reconstructed $c(z)$ vs. $c(z)=c_{0}$. Blue: residuals calculated from the best fit reconstructed values; red: residuals calculated from the $1\sigma$ lower confidence level from the reconstructed values; green: residuals calculated from the $2\sigma$ lower confidence level from the reconstructed values; yellow: residuals calculated from the $3\sigma$ lower confidence level from the reconstructed values. \label{fig:pprobabend}}
\end{figure*}

A more detailed inspection of the possibility to detect a VSL signal is given in Figure~\ref{fig:pprobabend}: for each simulation, and in each redshift bin, we calculate the residuals with respect to a constant speed of light, i.e. $c(z)/c_{0} = 1$; then, we plot the normalized number of simulations for which such residuals are positive, implying a clear detection of a non-constant $c(z)/c_{0}$. In blue, we show results when the residuals with respect to constant speed of light are calculated using the best fit Pad$\mathrm{\acute{e}}$ approximant, eq.~(\ref{eq:pade_cz}); in red, the residuals are calculated using the lower $1\sigma$ limit derived from the same best fit function, thus indicating a detection of the VSL signal at $1\sigma$ confidence level; in green and yellow, respectively, the residuals are calculated using the lower $2\sigma$ and $3\sigma$ limits. We focused on the lower limits because our fiducial input VSL corresponds to a speed of light higher than $c_{0}$ at higher redshift; in a more realistic case, one should check for both positive and negative residuals, and test, in the same way we are doing here, if there is any statistically clear evidence for one trend over the other.

Looking at Figure~\ref{fig:pprobabend} we can now have a more clear and precise prediction of what could happen in the next future. The probability to detect a $1\%$ VSL at a $3\sigma$ level is higher than the $95 \%$ (e.g., in $95 \%$ of our simulations we are able to detect a $1\%$ VSL signal at a $3\sigma$ level) approximately in the redshift range $[0.05,1.55]$. For higher redshifts, using the precision actually forecasted for next future galaxy surveys, we see that the signal degrades very rapidly. This point is interesting: given our input $c(z)$, the deviation from the $c(z)/c_{0} = 1$ limit grows with redshift. Thus, higher redshifts imply larger deviations. But this does not automatically converts in a clearer or easier detection: if the survey precision degrades too fast, we are going to lose any possibility to detect the signal at high redshift. At the same time, this is also encouraging: given the possibly larger deviation from the constancy of the speed of light at high redshifts (if the VSL signal is a monotonical function, of course), we have a lot of room to improve its detection in this range, because the precision update required for future surveys at higher redshifts is well inside our technological possibilities.

Another interesting point to stress is that in \citep{PRL15,BAOext} we have shown how SKA will be able to put a $3\sigma$ limit on a $1\%$ VSL signal at the maximum redshift in the angular diameter distance, which should locate at $z \sim 1.55-1.65$ (at least, basing this claim on our present knowledge of the cosmological background model). In the method exposed in this work, the detection at $3\sigma$ for a redshift $\approx 1.65$ is possible in the $80\%$ of our simulations; still a high probability, even if quite lower than the $95\%$ limit assessed above. It is not surprising that the two methods have different sensitivity at this redshift, because they rely on different algorithms; in particular, the maximum detection method described in \citep{PRL15,BAOext} can be pushed to a better precision, while the present method is mainly limited by the not-perfect correspondence of the derivatives calculated from real data with those intrinsic to the unknown cosmological background. But, still, the two methods are complementary, helping to extend the final redshift range of VSL detection.

Unfortunately, from Figure~\ref{fig:pprobabend} it is also clear that a $0.1\%$ signal will be hardly detected: at $1\sigma$ level, the probability detection of a VSL signal of such magnitude is $\sim 80 \%$ in the redshift range $[0.05,1.15]$; a $3\sigma$ detection in the same range is achieved only in $60 \%$ of our simulations, thus making it difficult to statistically state if it can be really reached or not.

\begin{figure*}[htbp]
\centering
  \includegraphics[width=0.9\textwidth]{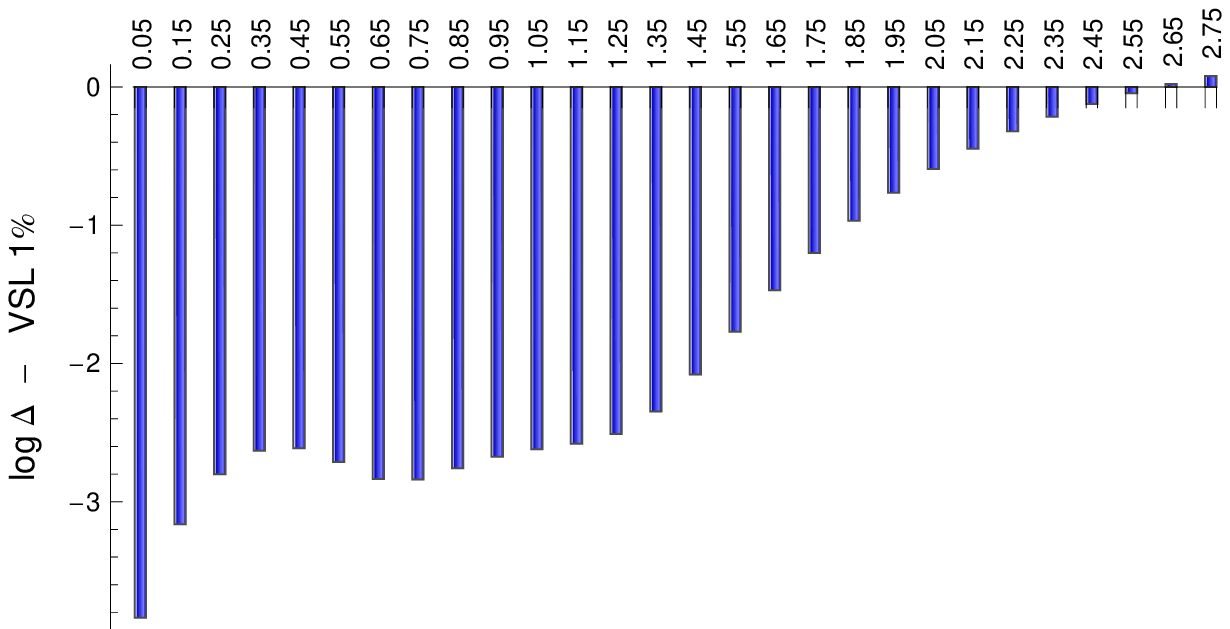}\\
  ~~~\\
  \includegraphics[width=0.9\textwidth]{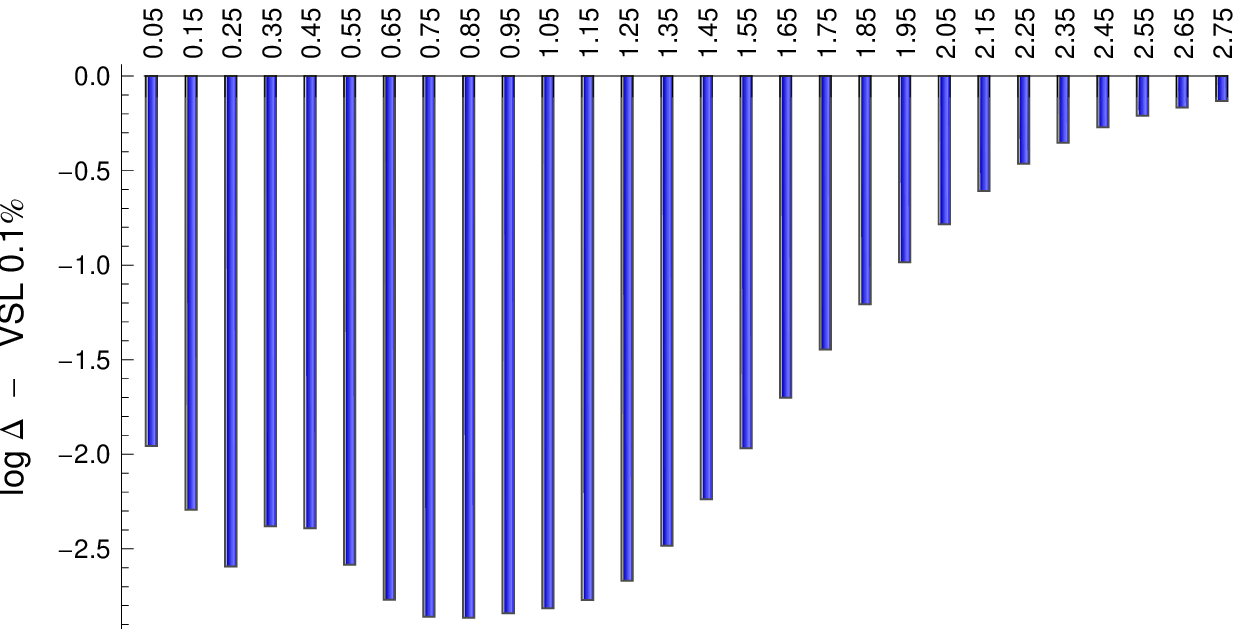}
  \caption{Reconstruction goodness criterium (residuals sum parameter) for the reconstruction of the underlying VSL signal. \label{fig:residual}}
\end{figure*}

Another important question to be stated is what level of goodness has our method to reconstruct the real VSL background. In order to assess this question, we calculate the quantity:
\begin{equation}
\Delta_{i} = \sum^{\mathcal{N}_{sim}}_{j=1} \frac{\left(c_{theo}(z_{i}) - c_{ans}(z_{i})\right)^2}{c^{2}_{ans}(z_{i})}\; ,
\end{equation}
where $\mathcal{N}_{sim} = 10^{3}$ is the total number of simulations we have run; $c_{theo}(z_{i})$ is the varying speed of light given by the resulting best Pad$\mathrm{\acute{e}}$ approximant, eq.~(\ref{eq:pade_cz}), evaluated at each redshift $z_{i}$; and $c_{ans}(z_{i})$ is the fiducial speed of light given by eq.~(\ref{eq:ansatz_c}). Thus, the quantity $\Delta_{i}$ is the sum, at each redshift and all over our simulations, of the relative squared residuals between our final reconstructed VSL $(c_{theo})$ and the fiducial one $(c_{ans})$. Smaller is its value, better is the agreement between our reconstruction and the true underlying model. In this way, we have a criterium to establish if our reconstruction is accurate or not. In Figure~\ref{fig:residual} we plot the logarithm of this quantity; conclusions are the same we have derived above: in the case of a $1\%$ VSL, it is clear that the agreement is quite good and very similar in the redshift range $[0.05,1.25]$; then it starts to decrease, with a $\Delta$ that, at $z \sim 1.65$, is one order of magnitude larger then the minimum value achieved; and things go even worse for larger redshift values. The same is more or less valid for the $0.1\%$ case.

\subsection{Curvature degeneracy}

As pointed out in previous sections, in the Friedmann and continuity equations the VSL-based terms, $c(t)$ and  $\dot{c}(t)$, come coupled with the spatial curvature parameter $(k)$. Thus, it is natural to expect some degree of degeneracy between a possible VSL signal and what could instead interpreted as a geometric effect. Our main equations are derived, as said previously, assuming that the Universe if spatially flat, i.e. $k=0$. All the most updated observations confirm such assumption \citep{PlanckCosmo}; but we want to show here that even if we take into account curvature, still there is a wide range of validity for our equations and, even, our method might be generalized and used, in the standard context of constant speed of light, as an alternative way to measure the spatial curvature.

\begin{figure*}[htbp]
\centering
  \includegraphics[width=0.75\textwidth]{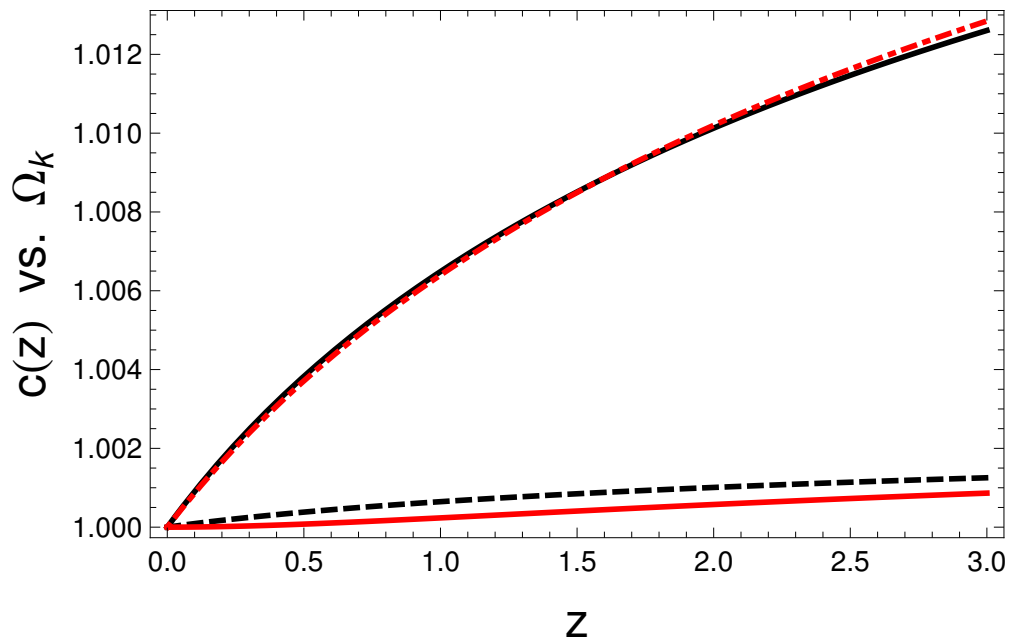}
  \caption{VSL vs. spatial curvature degeneracy displayed using eq.~(\ref{eq:rec_c_curv}). Black lines: solid - $1\%$ VSL signal (from eq.~(\ref{eq:ansatz_c}) plus null curvature ($\Omega_{k} = 0$ case in eq.~(\ref{eq:rec_c_curv})); dashed - $0.1\%$ VSL (from eq.~(\ref{eq:ansatz_c}) plus null curvature ($\Omega_{k} = 0$ case in eq.~(\ref{eq:rec_c_curv})). Red lines: correction from curvature term in eq.~(\ref{eq:rec_c_curv}) when $\Omega_{k}= 0.0008$ and assuming $c(z) = c_{0}$; dot-dashed - correction from curvature term in eq.~(\ref{eq:rec_c_curv}) when $\Omega_{k}= 0.0008$ and assuming a $0.95\%$ VSL signal. \label{fig:curvature}}
\end{figure*}

When taking into account spatial curvature, the main change is in the determination of what we have defined $y^{real}$, defined in eq.~(\ref{eq:h_real}) as the derivative of observational $D^{real}_{A}$ with respect to redshift. If the curvature is allowed to vary, then, the most general definition for the angular diameter distance is
\begin{equation}\label{eq:DA_k_VSL}
D_{A}(z) = \begin{cases}
\frac{D_{H}}{\sqrt{\Omega_{k}}(1+z)} \sinh \left( \frac{\sqrt{\Omega_{k}} D_{C}(z)}{D_{H}} \right) &\mbox{for } \Omega_{k} > 0\\
\frac{D_{C}(z)}{1+z} &\mbox{for } \Omega_{k} = 0 \\
\frac{D_{H}}{\sqrt{|\Omega_{k}|}(1+z)} \sin \left( \frac{\sqrt{|\Omega_{k}|} D_{C}(z)}{D_{H}} \right) &\mbox{for } \Omega_{k} < 0 \, ,
\end{cases}
\end{equation}
where $\Omega_k \equiv k c^{2}_{0} / H^{2}_{0}$ is the dimensionless curvature density parameter today; $D_{H} = c_{0} / H_{0}$ is the Hubble distance; and the line-of-sight comoving distance is defined as $D_{C}(z) = D_{H} \int^{z}_{0} \mathcal{F}_{c}(z') / E(z') dz'$, where we have made use of the general ansatz $c(z) \equiv c_{0} \mathcal{F}_{c}(z)$, with $\mathcal{F}_{c}(z) = 1$ for $z=0$. We are assuming here the most general case of a varying speed of light $c(z)$; but the standard scenario can be easily recovered simply replacing $c(z)$ with $c_{0}$ any time it appears. If we now calculate $y^{real}$ through the same eq.~(\ref{eq:h_real}), we have:
\begin{equation}\label{eq:rec_c_curv}
y^{real}_{r}(z) \equiv \begin{cases}
\frac{c(z)}{H(z)} \cosh \left( \frac{\sqrt{\Omega_{k}} D_{C}(z)}{D_{H}} \right) &\mbox{for } \Omega_{k} > 0\\
\frac{c(z)}{H(z)} &\mbox{for } \Omega_{k} = 0 \\
\frac{c(z)}{H(z)} \cos \left( \frac{\sqrt{|\Omega_{k}|} D_{C}(z)}{D_{H}} \right) &\mbox{for } \Omega_{k} < 0 \, .
\end{cases}
\end{equation}
It is clear that even if we assume $c(z) = c_{0}$, we would still have some contribution from the $\Omega_{k} \neq 0$ term; thus the case ``VSL $+$ spatial flatness'' would be equivalent to ``constant $c(z)$ $+$ curvature''. We can easily quantify how much information we might derive, and which we might erroneously attribute to a VSL signal only, should instead be shared with a non-null curvature signal. From the \emph{Planck} Legacy Archive, the extension of the baseline model with a free curvature parameter, named \texttt{base}$\_$\texttt{omegak}$\_$\texttt{plikHM}$\_$\texttt{TTTEEE}$\_$\texttt{lowTEB}$\_$\texttt{BAO}$\_$\texttt{H070p6}$\_$\texttt{JLA}$\_$\\
\texttt{post}$\_$\texttt{lensing}, gives the value of $\Omega_k = 0.0008 \pm 0.002$ at the $68\%$ confidence level (and $\pm 0.004$ at the $95\%$). We can thus compare the curvature-correction terms in eq.~(\ref{eq:rec_c_curv}) with the null curvature hypothesis, using our ansatz for the VSL, eq.~(\ref{eq:ansatz_c}). Results are shown in Figure~\ref{fig:curvature}.

The first possible conclusion is that a realistic contribution from the spatial curvature to our method (red line) would be $\sim 0.05\%$ at the maximum in $D_{A}$ (for a more direct and straightforward comparison, we use the same maximum criterium we have used to define the $1\%$ and the $0.1\%$ VSL models); thus, it would be even smaller than the $0.1\%$ VSL signal (black dashed line) we have considered so far, and would thus result, finally, undetectable. This result, obtained in an independent and alternative way, is also consistent with a recent attempt described in \citep{BAOK}.

For the sake of precision, we have to stress again that, anyway, in general, a pure VSL and a pure curvature signal are degenerate. We can detect a total signal, without being able to ascribe it to one or another. What we can estimate is that, given present bounds on curvature, a $1\%$ signal (solid black line) could be attributed with no doubt to VSL only, rather than to any curvature contribution. Even in the case of assuming both a VSL and non-null curvature (dot-dashed redline), given the actual constraints on the latter one, the VSL signal might be $\sim0.95\%$, in order to have a final total $1\%$ detection. Thus, at least at the scales which we have shown to be directly testable in the next future, curvature might play a negligible role. But if the total signal should result to be less than $1\%$, then we could have problems and would not be able to discriminate between them.

\section{Conclusions}
\label{sec:conclusions}

In this work we have extended the method previously described in \citep{PRL15,BAOext}: while the latter made it possible to measure the speed of light (and, incidentally, detect any possible variation of the same quantity) only in one well-located point (the maximum redshift), here we show how it is possible to recover a \emph{redhisft-extended} VSL signal on a much large redshift range. We have used the cosmological observations which will be available in the next future from galaxy surveys, i.e.: estimations of the sound horizon at decoupling/dragging epoch, imprinted as an angular diameter distance $D_{A}$ in the clustering of the galaxies; and the expansion rate data $H$ inferred from ETG galaxies designated as cosmological clocks. We have employed a quite various number of future galaxy surveys, BOSS, DESI, \emph{WFirst-2.4} and SKA, which result to have the best performances in different (non-overlapping) redshift windows.

As we have discussed in section~(\ref{sec:application}), and as it is shown in figure~\ref{fig:pprobabend}, given the sensitivities forecast for the previous surveys, there is a quite high probability $(>95\%)$ to detect a $1\%$ VSL signal (if any) at $3\sigma$ confidence level in the redshift range $z \in [0., 1.55]$. Smaller signals, of the order of $0.1\%$, will be hardly detected by the same surveys.

We have also given a more detailed discussion about the impact that a possible non-null spatial curvature might have on the detection of the VSL signal. In particular, we have shown that values of the curvature compatible with the present bounds given by \emph{Planck} are absolutely negligible with respect to a $1\%$ VSL signal. We emphasize here that, even if we were not considering a VSL signal, but the classical constancy of the speed of light, then our method would result to be useful to detect \emph{curvature-only} contributions. In particular, if such contribution should result to be $\approx 0.01$, then, it would be equivalent to a $1\%$ VSL signal, and all the discussion we have spent for the VSL theory might be equivalently exported to spatial curvature measurements only.

More problematic would be to disentangle smaller VSL signals, which would result to be of the same order of the geometrical contribution; but, as we have shown here, such small signals are out of the detection possibilities of currently forecast galaxy surveys for the next $15$ years. In the meantime, we may work to improve our method, and/or find alternative ones.

\section*{Acknowledgments}

This work is financed by the Polish National Science Center Grant DEC-2012/06/A/ST2/00395.

\vfill
\end{document}